\documentclass[twocolumn,showkeys,aps,prb,showpacs]{revtex4-1}
\usepackage{graphicx}
\UseRawInputEncoding
\usepackage[CJKbookmarks,dvipdfm,colorlinks,linkcolor=blue,citecolor=blue]{hyperref}
\usepackage{amsmath, amssymb}
\usepackage{makecell}
\usepackage{multirow}
\usepackage{CJK}
\usepackage{mathrsfs}
\usepackage{bm}
\usepackage{amsmath}
\usepackage{dcolumn}
\usepackage{epstopdf}
\usepackage{dsfont}
\usepackage{amssymb}
\usepackage{tabularx}
\usepackage{array}
\usepackage{float}
\usepackage{color}
\usepackage{epstopdf}
\usepackage{mathrsfs}
\usepackage{extarrows}

\begin{document}

\title{ Intrinsic persistent spin helix  in 2D T-XY (X$\neq$Y=P, As, Sb and Bi)}

\author{San-Dong Guo$^{1}$, Xu-Kun Feng$^{2}$,  Dong Huang$^1$, Shaobo Chen$^3$ and Yee Sin Ang$^{2}$}
\affiliation{$^1$School of Electronic Engineering, Xi'an University of Posts and Telecommunications, Xi'an 710121, China}
\affiliation{$^2$Science, Mathematics and Technology (SMT), Singapore University of Technology and Design (SUTD), 8 Somapah Road, Singapore 487372, Singapore}
\affiliation{$^3$College of Electronic and Information Engineering, Anshun University, Anshun 561000, People's Republic of China}
\begin{abstract}
The persistent spin helix (PSH) is robust against spin-independent scattering and renders an
extremely long spin lifetime, which can improve the
performance of potential spintronic devices.   To achieve the PSH,   a
unidirectional spin configuration is required in the momentum space. Here, T-XY (X$\neq$Y=P, As, Sb and Bi) monolayers  with dynamical, mechanical and thermal  stabilities  are predicted to intrinsically possess   PSH.  Due to the $C_{2\upsilon}$ point-group symmetry, a unidirectional spin configuration is preserved in the out-of-plane direction for both conduction and valence bands around the high-symmetry $\Gamma$ point. That is,  the expectation value of the spin
$S$ only has the out-of-plane component $S_z$.  The application of an out-of-plane external
electric field can induce in-plane components $S_x$ and $S_y$, thus offering a promising platform for the on-off logical functionality of spin devices.
 Our work reveals a new family of T-phase two-dimensional (2D) materials, which could provide promising applications in spintronic  devices.

\end{abstract}
\keywords{Spin-orbital coupling, Persistent spin helix~~~~~~~~~~~~~~~~~~~~~~Email:sandongyuwang@163.com}

\maketitle

\section{Introduction}
 In a  material  with  broken inversion symmetry, the spin-orbital coupling (SOC) induces momentum ($k$)-dependent spin-orbit field (SOF), which can lift spin degeneracy and leads to the
nontrivial $k$-dependent spin textures of the spin-split bands through  Rashba  and Dresselhaus  effects\cite{p1,p2}.
Practically, the strong
Rashba SOC  allows for electrostatic manipulation of the spin states, which has potential application for  non-charge-based computing and information processing\cite{p3,p4,p5}.
However, strong SOC can cause  spin decoherence, which leads to the reduced spin lifetime.
The impurities and defects can change the momentum of electrons in a diffusive transport regime, and simultaneously
randomize the spin due to the $k$-dependent SOF, which induces  spin decoherence through the Dyakonov-Perel (DP) mechanism and limits the
performance of potential spintronic devices\cite{p6}.

To overcome spin dephasing, a possible way  is designing
a structure with the SOF orientation to be
unidirectional,  which can preserve  a unidirectional spin configuration
in the $k$ space.
The unidirectional SOF will
lead to a spatially periodic mode of the spin polarization, which is  known as the persistent spin helix (PSH)\cite{p7,p8}. The spin dephasing  can be
suppressed by PSH  due to SU(2) spin rotation symmetry, which renders an
extremely long spin lifetime\cite{p7,p9}. The PSH
has been demonstrated on various  quantum
well (QW) heterostructures, interface and  surface\cite{p10,p11,p12,p13,p13-1,p14,p15}. Recently, a different approach for achieving the PSH is imposing
the specific symmetry of the crystal, which has been realized in  bulk-, layered-   and  two-dimensional (2D)-ferroelectric systems\cite{p16,p17,p18,p19,p20,p21,p22,p23,p24,p25,p26}.
However, the search for 2D materials with intrinsic PSH has lately been very
demanding of attention due to their potential for miniaturization spintronic devices.

\begin{figure*}
  % Requires \usepackage{graphicx}
  \includegraphics[width=15cm]{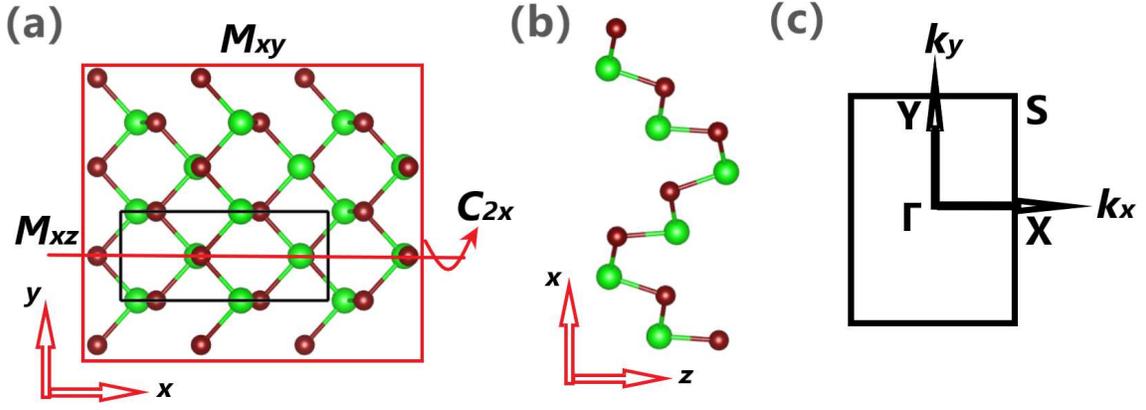}
  \caption{(Color online)For monolayer T-XY (X$\neq$Y=P, As, Sb and Bi),  the top view (a) and  side view (b) of  crystal structure with large balls for X atoms and small balls for Y atoms, and  the related symmetry operations are shown in (a). The (c)  shows the first BZ with high-symmetry points. }\label{t0}
\end{figure*}

\begin{figure}
  % Requires \usepackage{graphicx}
  \includegraphics[width=7cm]{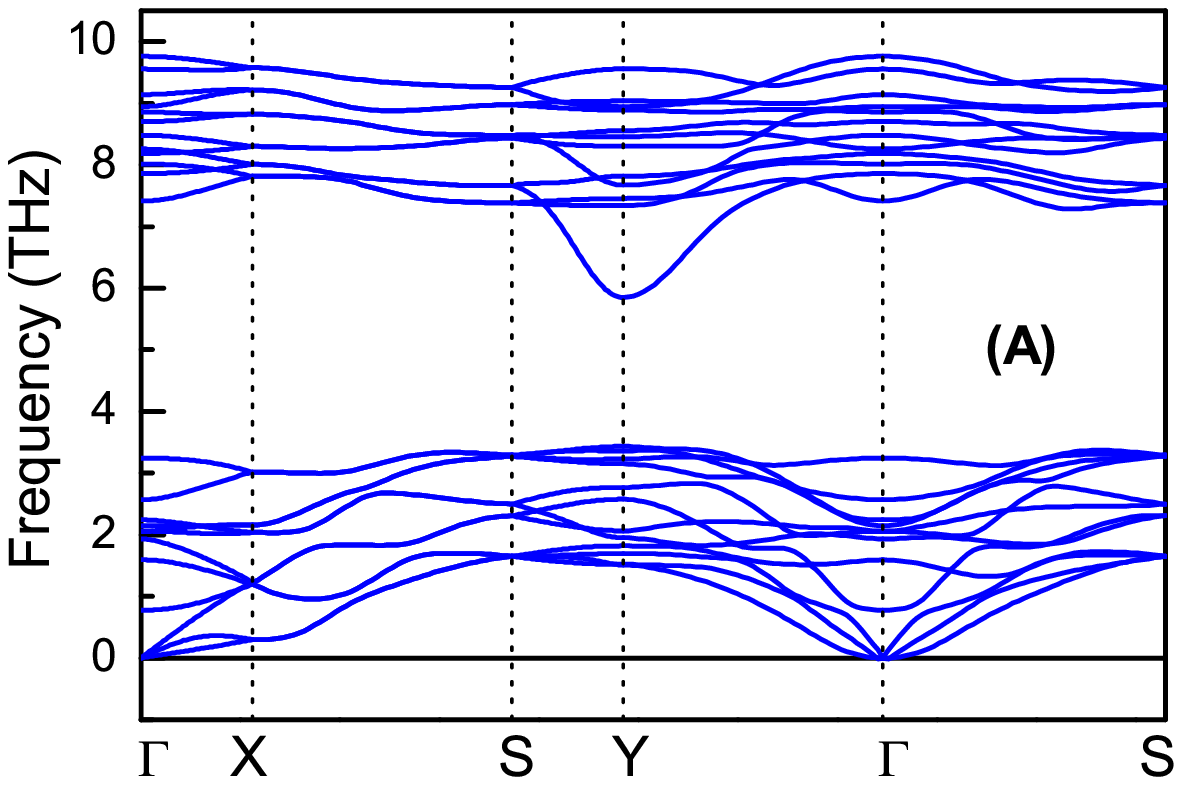}
  \includegraphics[width=7.5cm]{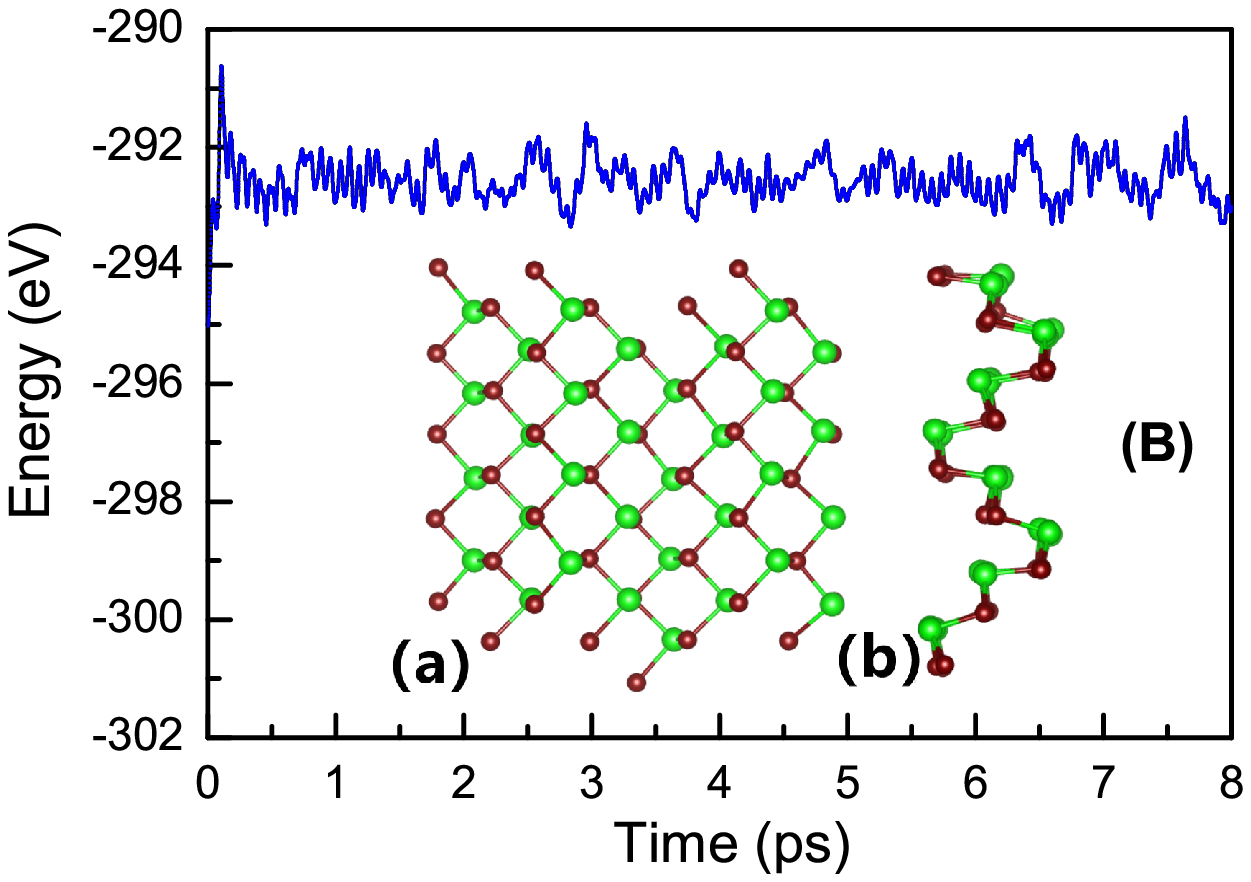}
  \caption{(Color online)For T-SbP monolayer, (A): the phonon dispersion curves;  (B):  the variation of free energy during the AIMD simulation. Insets show the
 final structures (top view (a) and side view (b))  after 8 ps at 300 K. }\label{t1}
\end{figure}

Recently, similar to the $\alpha$-phase, the T-phase MZ (M=Sn and  Ge; Z=S and Se) has been proposed with excellent thermal, dynamical and mechanical stabilities, which exhibits $C_{2\nu}$ symmetry and fold
characteristics similar to "triple-staggered layers"\cite{apl}.
These T-MX monolayers possess fine piezoelectric
performance and stability\cite{apl}. Here,  the T-XY (X$\neq$Y=P, As, Sb and Bi) monolayers are proposed with dynamical, mechanical and thermal  stabilities, which possess the same  number of valence electrons  with  MX (M=Sn and Ge; X=S and Se).
The T-XY are predicted to intrinsically possess   PSH with a unidirectional out-of-plane spin configuration ($S_z$)  for both conduction and valence bands due to the $C_{2\upsilon}$ point-group symmetry. When an out-of-plane external
electric field is applied,  the in-plane components $S_x$ and $S_y$ can be induced, which  offers  a promising platform for the on-off logical functionality of spin devices.

The rest of the paper is organized as follows. In the next
section, we shall give our computational details and methods.
 In  the next few sections,  we shall present crystal structure and stability, electronic structures and PSH of T-XY (X$\neq$Y=P, As, Sb and Bi) monolayers. Finally, we shall give our discussion and conclusion.

\section{Computational detail}
 Within density functional theory (DFT)\cite{1}, the first-principle calculations are carried out  by using the projector augmented wave (PAW) method as implemented in Vienna ab initio Simulation Package (VASP)\cite{pv1,pv2,pv3}. We use the generalized gradient approximation  of Perdew, Burke and  Ernzerhof  (GGA-PBE)\cite{pbe}  as  exchange-correlation  functional.
The SOC is included to investigate electronic structures and PSH of T-XY (X$\neq$Y=P, As, Sb and Bi), and the details of the SOC
implementation in the PAW methodology are given in ref.\cite{soc}.
The energy cut-off of 500 eV, total energy  convergence criterion of  $10^{-7}$ eV and force
convergence criteria  of  0.001 $\mathrm{eV.{\AA}^{-1}}$ are set to perform the first-principles calculations.
A vacuum space of
more than 28 $\mathrm{{\AA}}$ between slabs along the $z$ direction is added to eliminate the spurious
interactions.
A 3$\times$5$\times$1 supercell is used to calculate the phonon spectrum within the finite displacement method by using the  Phonopy code\cite{pv5}.
The elastic  stiffness ($C_{ij}$) are obtained by strain-stress relationship (SSR), and the 2D  $C^{2D}_{ij}$
have been renormalized by   $C^{2D}_{ij}$=$L_z$$C^{3D}_{ij}$, where  $L_z$ is  the length of unit cell along $z$ direction.
The constant energy contour plots of the spin
texture are calculated by the PYPROCAR code\cite{py}.
A 9$\times$18$\times$1 k-point meshes in the first
Brillouin zone (BZ) are adopted  for all calculations.

\begin{table*}
\centering \caption{For T-XY (X$\neq$Y=P, As, Sb and Bi) monolayer, the lattice constants $a_0$ and $b_0$ ($\mathrm{{\AA}}$),   the elastic constants $C_{ij}$ ($\mathrm{Nm^{-1}}$), the  Young's modulus along $x$ and $y$ directions $C_\text{2D}(x)$ and $C_\text{2D}(y)$ ($\mathrm{Nm^{-1}}$), and the GGA and GGA+SOC gaps $Gap$ and $Gap^\text{SOC}$ (eV). }\label{tab0}
  \begin{tabular*}{0.96\textwidth}{@{\extracolsep{\fill}}ccccccccccc}
  \hline\hline
 Name &$a_0$&$b_0$ &$C_{11}$& $C_{12}$&$C_{22}$&$C_{66}$&$C_\text{2D}(x)$&$C_\text{2D}(y)$&$Gap$&$Gap^\text{SOC}$  \\\hline\hline
$\mathrm{AsP}$ &   9.38  &3.48   & 9.87 & 10.72& 77.64&18.97 &8.39&66.00 &1.050&1.047              \\\hline
$\mathrm{SbAs}$ &  9.66& 3.94&7.12&11.02&55.07&13.33&4.92&38.01 &0.307&0.289\\\hline
$\mathrm{SbP}$ &   8.95& 3.83& 3.50&9.19&60.04&17.71&2.09&35.91 &0.366&0.332 \\\hline
$\mathrm{BiP}$ &   8.40& 4.00& 4.83&3.34&54.15&17.10&4.62&51.84 &0.621&0.381 \\\hline
$\mathrm{BiAs}$ &   9.45& 4.10& 7.99&9.49&49.51&12.19&6.17&38.24&0.319&0.108  \\\hline
$\mathrm{BiSb}$ &   9.94& 4.35& 7.49&9.94&43.04&10.64 &5.19&29.85&0.354&0.026 \\\hline\hline
\end{tabular*}
\end{table*}

\section{Crystal structure and stability}
 The top and side views of  crystal structures of T-XY (X$\neq$Y=P, As, Sb and Bi) are shown in
\autoref{t0} (a) and (b).  The primitive cell  contains eight atoms,
including four X atoms  and four Y atoms, and each X (Y)
atom is connected to three surrounding Y (X) atoms.  Their optimized lattice parameters $a$ and $b$ along
the $x$ and $y$ directions are listed in \autoref{tab0}.

The T-phase has four symmetry operations: (i)
identity operation $E$; (ii) twofold screw rotation $\bar{C}_{2x}$; (iii) glide reflection $\bar{M}_{xy}$
; and (iv) reflection $M_{xz}$ with respect to the $xz$ plane.
The $\bar{C}_{2x}$ is performed by
twofold rotation around the x axis ($C_{2x}$), followed by translation of
$\tau$=($a$/2, $b$/2). The  $\bar{M}_{xy}$ can be obtained by reflection with respect to the $xy$ plane followed by translation $\tau$. The T-phase lacks  inversion symmetry,  allowing piezoelectric response. Because these T-XY monolayers share similar behavior, in the following, we shall mainly focus on T-SbP monolayer as a representative. The results of the remaining five monolayers are briefly mentioned or put in FIG.1, FIG.2, FIG.3, FIG.4, FIG.5, FIG.6 and FIG.7\cite{bc}.
\begin{figure}
  % Requires \usepackage{graphicx}
     \includegraphics[width=8cm]{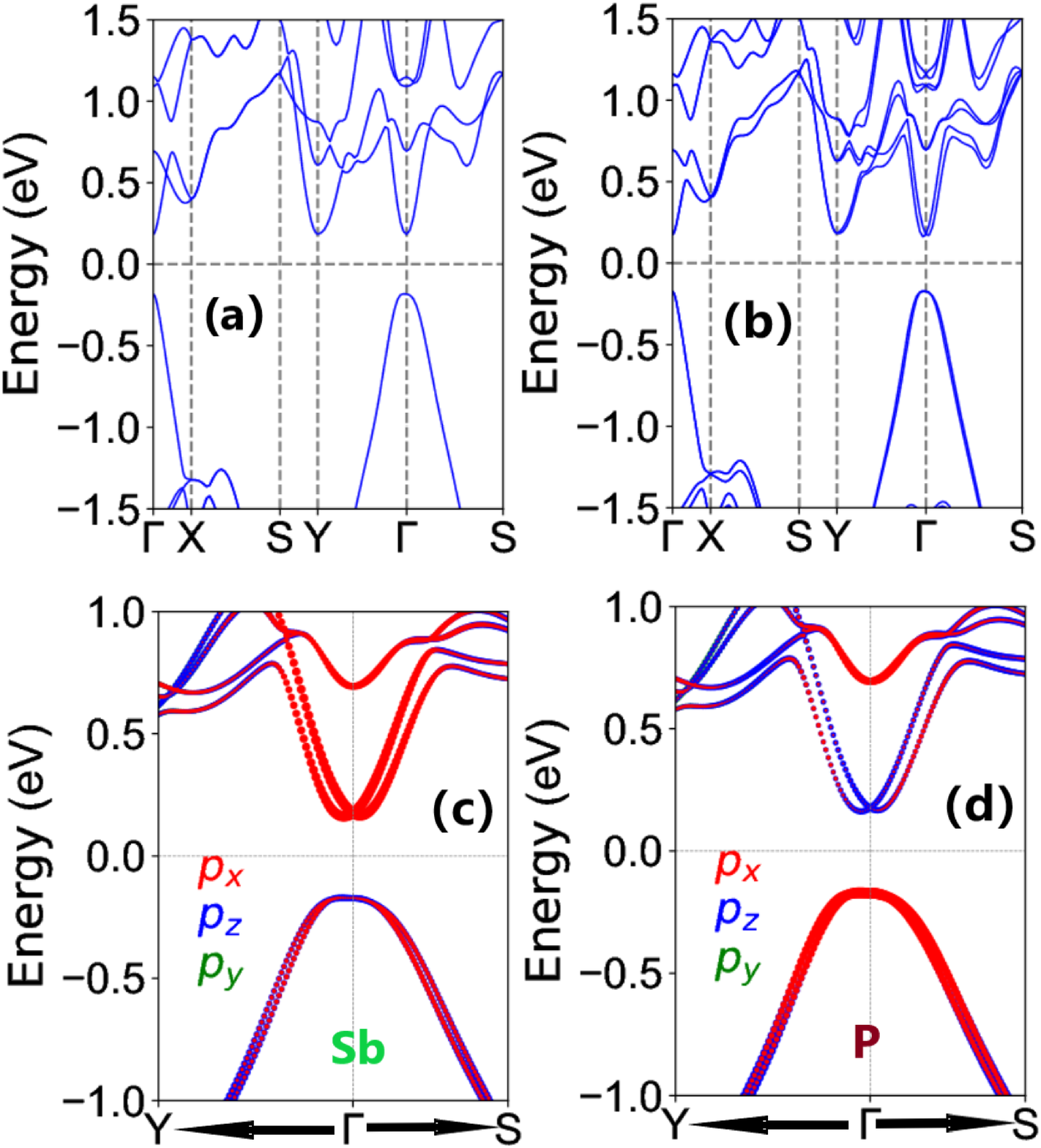}
  \caption{(Color online)For T-SbP, the energy band structures without SOC (a) and with SOC (b).  Zoom-in on the energy band structures around
   $\Gamma$ point along $\Gamma$-Y and $\Gamma$-S lines with projected Sb (c) and P (d) atomic $p$-orbitals.}\label{t2}
\end{figure}

To  verify their dynamical stabilities, their phonon dispersions are calculated.  \autoref{t1} (A) plots the result for monolayer
SbP, which shows  no imaginary frequency, indicating that the structure is dynamically stable.
To investigate the thermal stability, the  ab-initio molecular dynamics (AIMD) simulations are performed with a 2$\times$4$\times$1 supercell and a time step
of 1 fs. \autoref{t1} (B) shows the simulation result on monolayer T-SbP at 300 K for 8 ps. One observes that the
energy  fluctuates  within a small range during the whole simulation time, and the overall structure is well maintained at this temperature, indicating its good thermal stability.
The linear elastic constants  are calculated to determine their mechanical stabilities.
Due to  $C_{2\upsilon}$ symmetry, four independent elastic constants ($C_{11}$, $C_{12}$, $C_{22}$ and $C_{66}$) can be observed, which are listed in \autoref{tab0}.
These $C_{ij}$ meet Born-Huang
criteria of  mechanical stability  ($C_{11}>0$, $C_{66}>0$ and  $C_{11}*C_{22}>C_{12}^2$)\cite{ela},  thereby verifying their mechanical
stabilities. The direction-dependent Young's modulus $C_\text{2D}(\theta)$ can be obtained as\cite{ela1}:
\begin{equation}\label{c2d}
C_\text{2D}(\theta)=\frac{C_{11}C_{22}-C_{12}^2}{C_{11}\sin^4\theta+A\sin^2\theta \cos^2\theta+C_{22}\cos^4\theta},
\end{equation}
where $\theta$ is the polar angle measured from $x$, and $A=(C_{11}C_{22}-C_{12}^2)/C_{66}-2C_{12}$.
According to FIG.8 and FIG.9\cite{bc},  obvious anisotropy of elasticity is  observed in these T-XY monolayers.
The Young's modulus of T-XY along $x$ and $y$ directions are listed \autoref{tab0}. It is clearly seen that $C_\text{2D}(x)$
 is very smaller than $C_\text{2D}(y)$. This indicates that the T-XY is more flexible and deformable along $x$ direction, which plays a positive
effect on their piezoelectricity\cite{apl}.

Although these T-XY monolayers  have not been experimentally synthesized,  the above calculated  results show that they
possess  good thermal stability  as well as positive dynamical and mechanical stabilities. Therefore, we expect that these monolayers can be synthesized experimentally in the future.

\begin{figure*}
  % Requires \usepackage{graphicx}
     \includegraphics[width=14cm]{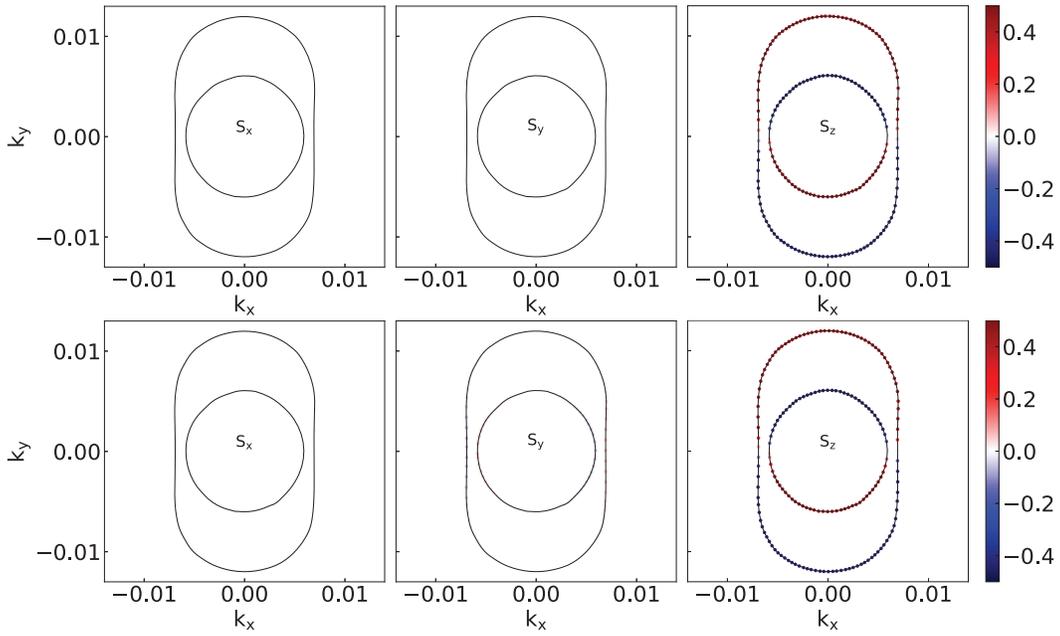}
  \caption{(Color online)For T-SbP monolayer, the top plane/bottom plane shows the spin texture ($S_x$, $S_y$ and $S_z$) calculated in a $k_x-k_y$ plane centered at the $\Gamma$ point with the isoenergetic
 surface of 0.25  eV  above  the Fermi level at $E$=0.00/0.30 $\mathrm{V/{\AA}}$.  The color scale  show the
modulus of the spin polarization.  Note: the picture should be enlarged enough to see the details.}\label{t3}
\end{figure*}
\begin{figure*}
  % Requires \usepackage{graphicx}
     \includegraphics[width=14cm]{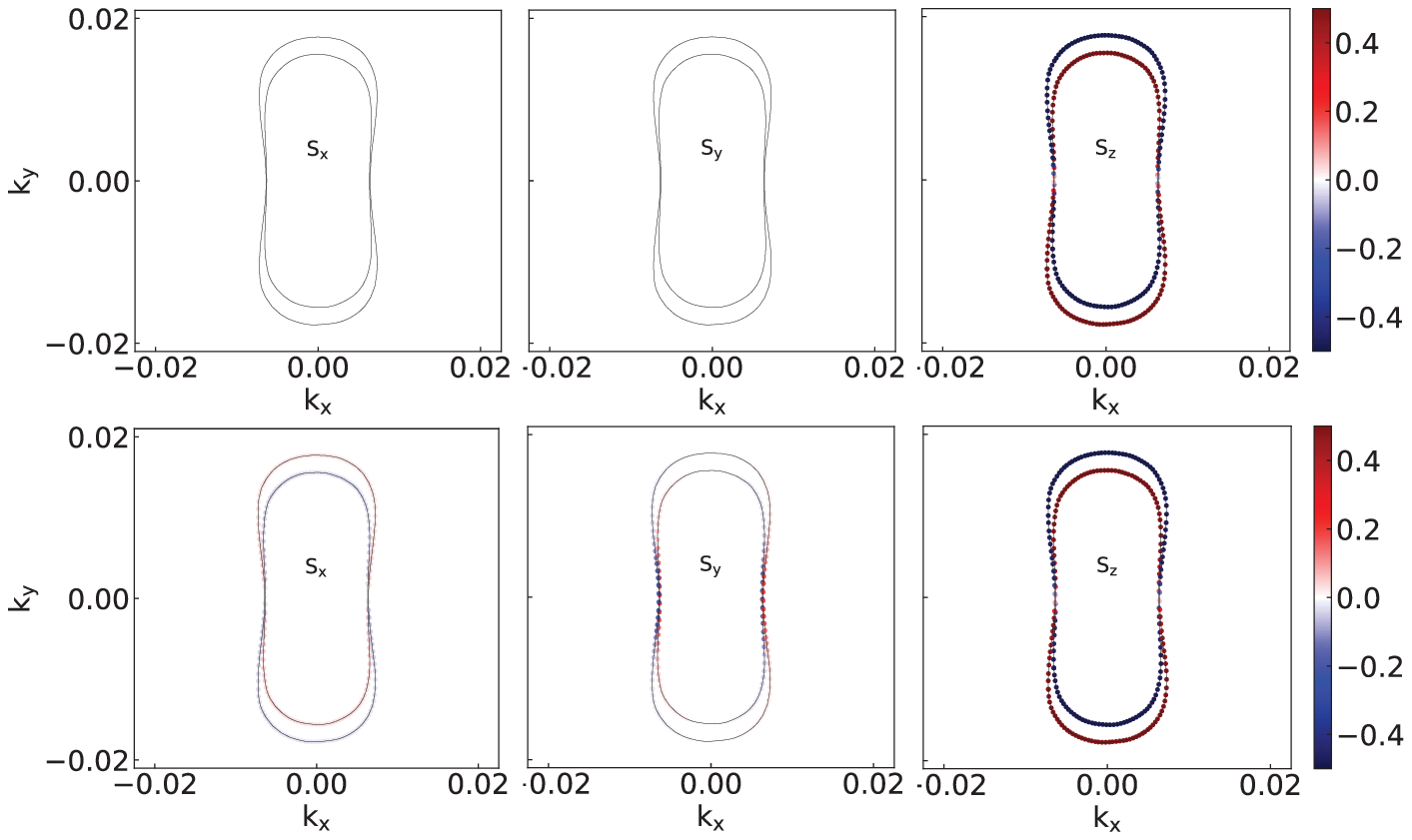}
  \caption{(Color online)For T-SbP monolayer, the top plane/bottom plane shows the spin texture ($S_x$, $S_y$ and $S_z$) calculated in a $k_x-k_y$ plane centered at the $\Gamma$ point with the isoenergetic
 surface of -0.25  eV  below  the Fermi level at $E$=0.00/0.30 $\mathrm{V/{\AA}}$.  The color scale  show the
modulus of the spin polarization.  Note: the picture should be enlarged enough to see the details.}\label{t5}
\end{figure*}
\begin{figure}
  % Requires \usepackage{graphicx}
     \includegraphics[width=8cm]{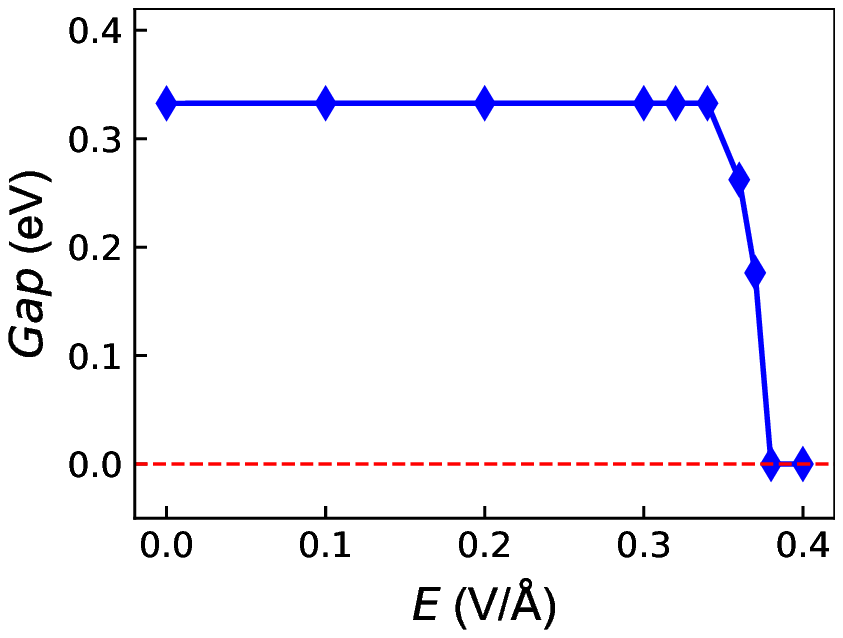}
  \caption{(Color online)For T-SbP monolayer, the energy band gap as a function of out-of-plane external electric field $E$.}\label{t4}
\end{figure}

\section{electronic structures and persistent spin helix}
\autoref{t2} shows the energy  band structures of T-SbP calculated
along high-symmetry $k$  paths, and those of  the other five are plotted in FIG.3, FIG.4, FIG.5, FIG.6 and FIG.7\cite{bc}. Without including the SOC, the T-SbP, T-BiP, T-BiAs and T-BiSb are  indirect bandgap semiconductors, and the  conduction band bottom (CBM)  locates at $\Gamma$ point, while the valence band maximum (VBM) slightly deviates from $\Gamma$ along the $\Gamma$-Y path. However, the T-AsP and T-SbAs are direct bandgap semiconductors with both CBM and VBM at $\Gamma$ point. The gaps of T-AsP, T-SbP, T-SbAs, T-BiP, T-BiAs and T-BiSb are 1.050 eV, 0.366 eV, 0.307 eV, 0.621 eV, 0.319 eV and 0.354 eV, respectively.
When including the SOC,  the energy band
structures of T-XY are modified, and they all become indirect bandgap semiconductors with both CBM and VBM slightly deviating from $\Gamma$ point.
The GGA+SOC  gaps of six monolayers are reduced with respect to GGA ones.  For  T-XY (X$\neq$Y=P, As and Sb), the changes of gap are small, while the gaps of T-BiY (Y=P, As and Sb) vary greatly.
The gaps of T-AsP, T-SbP, T-SbAs, T-BiP, T-BiAs and T-BiSb become  1.047 eV, 0.332 eV, 0.289 eV, 0.381 eV, 0.108 eV and 0.026 eV, respectively. The related gap data are summarized in \autoref{tab0}.
Importantly, a
sizable splitting of the bands produced by the SOC is observed. This splitting is especially pronounced around the $\Gamma$ point near both the VBM and CBM. According to \autoref{t2} (c) and (d),  the Sb-$p_x$
and P-$p_z$ orbitals contribute dominantly to the CBM, while the
VBM  are mainly  from the contributions of the Sb-$p_z$ and
P-$p_x$ orbitals.

For T-SbP monolayer, the top planes of  \autoref{t3} and \autoref{t5} show the spin textures of conduction and valence bands calculated in a $k_x-k_y$ plane centered at the $\Gamma$ point with the isoenergetic
 surface of 0.25 eV  and -0.25 eV, respectively.  It is clearly seen that the spin polarization is originated from the out-of-plane component $S_z$, while the in-plane components of spin ($S_x$ and  $S_y$) are  zero. This means that the spin texture is  unidirectional, which is very different from the in-plane Rashba spin textures. Such a spin texture produces a unidirectional out-of-plane SOF, resulting in a persistent spin texture (PST).  The unidirectional SOF is robust against spin-independent scattering due to the PSH state, which leads to an extremely long spin lifetime by suppressing the DP spin-relaxation mechanism\cite{p6}.

To explain the PST with  only  out-of-plane component $S_z$, we establish an effective $k\cdot p$ Hamiltonian expanded at $\Gamma$ point. The twofold degeneracy at $\Gamma$ corresponds to the $\Gamma_{5}$ double-valued irreducible representation of $C_{2v}$. Taking the two states as basis, the generators, $C_{2x}$ and $M_{xy}$,  are represented in the matrix form as
\begin{equation}
\begin{split}
    D(C_{2x}) = -i\sigma_{x}\\
    D(M_{xy}) = -i\sigma_{z}\\
 D(\mathcal{T}) = -i\sigma_{y}\mathcal{K}
 \end{split}
 \end{equation}
 $\mathcal{T}$ is time reversal symmetry operation. These symmetries constrain the Hamiltonian $H$ by
\begin{equation}
\begin{split}
    {C}_{2x}H(\boldsymbol{k}){C}_{2x}^{-1} = H(k_{x}, -k_{y})\\
%\end{equation}
%\begin{equation}
    {M}_{xy}H(\boldsymbol{k}){M}_{xy}^{-1} = H(k_{x}, k_{y})\\
%\end{equation}
%\begin{equation}
    \mathcal{T}H(\boldsymbol{k})\mathcal{T}^{-1} = H(-k_{x}, -k_{y})
    \end{split}
\end{equation}
Expanding $H(k)$ to $k^{3}$ order, the Hamiltonian can be written as:
\begin{equation}\label{h}
    H(k) = \alpha k_{y}\sigma_{z}+\alpha_{1}k^{2}\sigma_{0}+(\alpha_{2}k_{y}^{3}+\alpha_{3}k_{x}^{2}k_{y})\sigma_{z}
\end{equation}
where $\sigma_{0}$ is 2$\times$2 identity matrix, $\sigma_{z}$ is the Pauli matrices, $\vec{k}$ ($k_x$ and $k_y$) is the wave vector, and   $\alpha_{i}$ are model parameters. Since the  spin operator $S_z$ commutes
with the Hamiltonian \autoref{h}, the spin operator $S_z$ is
a conserved quantity.  The expectation value of the spin $S$ only has the out-of-plane component, which produces the unidirectional out-of-plane spin configuration in $k$ space.

On the other
hand, by  ignoring the high-order $k$ items,
the linear-term parameter $\alpha$,  through the relation $\alpha=2E_R/k_0$,  can be obtained,
where $E_R$ and $k_0$ are the shifting energy and the wave vector along the $y$ direction.
The formation
of the PSH mode should have a substantially small wavelength $\lambda$
of the spin polarization. Here, the wavelength $\lambda$
can be estimated by using $\lambda=\frac{\pi\hbar^2}{\alpha m^*}$, where $m^*$ is the electron effective mass (Here, we only consider the conduction bands due to large spin-splitting.).
The $m^*$ is evaluated from the band
dispersion along the $\Gamma$-Y line in the CBM. For several selected 2D PSH systems, the spin-splitting parameter $\alpha$,  effective mass $m^*$,  and the wavelength of the spin polarization $\lambda$ are summarized in \autoref{tab3}.
 These $\alpha$ of T-SbP, T-BiP, T-BiAs and T-BiSb are comparable with those reported for several 2D  PSH systems, and the small wavelength of
the PSH mode is typically
on the scale of the lithographic dimension used in the recent
semiconductor industry. Thus, the T-SbP, T-BiP, T-BiAs and T-BiSb should be promising for miniaturization of spintronic devices.

\begin{table}
\centering \caption{For several selected 2D PSH systems [T-XY (X$\neq$Y=P, As, Sb and Bi), MN (M=Sn and Ge; Y=S, Se and Te) and GaMN (M=Se and Te; N=Cl, Br and I)],  the spin-splitting parameter $\alpha$ (eV$\cdot$\AA),  the effective mass $m^*$ ($m_0$),  and the wavelength of the spin polarization $\lambda$ (nm).}\label{tab3}
\begin{tabular*}{0.48\textwidth}{@{\extracolsep{\fill}}ccccc}
  \hline\hline
Monolayer & $\alpha$&	$m^*$&	$\lambda$& Reference	\\\hline\hline
AsP&0.059&0.120 &67.675& This work\\
SbAs&0.454&0.092&11.471&This work\\
SbP&1.103&0.116&3.745&This work\\
BiP&2.454&0.094&2.077&This work\\
BiAs&4.359&0.067&1.640& This work\\
BiSb&7.113&0.056&1.203&This work\\\hline\hline
MN &0.07$\sim$1.67& - &1.82$\sim$890&Ref.\cite{p24}\\
GaMN& 0.53$\sim$2.65 &-&1.20$\sim$6.57&  Ref.\cite{p20}\\\hline\hline
\end{tabular*}
\end{table}

The application of the external electric field can  modify the spin texture  of T-SbP. So,  an external out-of-plane electric field  $E$ is applied, which breaks both the $C_{2x}$ rotational and
$M_{xy}$ in-plane mirror symmetries. The gap of T-SbP as a function of $E$ is plotted in \autoref{t4}, and the related energy band structures are shown in FIG.10\cite{bc}. It is clearly seen that the gap basically remains unchanged, when $E$ is less than about 0.34  $\mathrm{V/{\AA}}$. At about $E$=0.38  $\mathrm{V/{\AA}}$, a semiconductor-to-metal phase transition can be observed.

Under the out-of-plane external electric field, the effective Rashba contribution  should be added in Hamiltonian \autoref{h}, which has the
isotropic form\cite{p1}:
\begin{equation}\label{a}
H^{iso}_R=\alpha^{iso}_R(k_x\sigma_y-k_y\sigma_x)
\end{equation}
The introducing Rashba term will induce
the in-plane spin components due to the broken in-plane mirror symmetry $M_{xy }$ caused by the  out-of-plane external electric field.
 To confirm this, the bottom planes of  \autoref{t3} and \autoref{t5} show the spin textures of conduction and valence bands calculated in a $k_x-k_y$ plane centered at the $\Gamma$ point with the isoenergetic
 surface of 0.25 eV and -0.25 eV  for T-SbP  at $E$=0.30  $\mathrm{V/{\AA}}$.
 It is clearly observed that significant in-plane spin components ($S_x$ and  $S_y$) are induced,  thus breaking the PST and perturbing the PSH state.

\begin{figure}
  % Requires \usepackage{graphicx}
  \includegraphics[width=8cm]{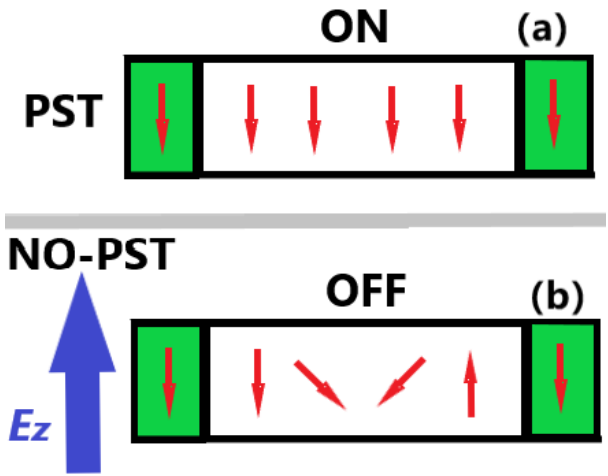}
  \caption{(Color online) Schematic of spintronic devices: the red arrows represent spin, while the blue arrows represent out-of-plane electric field.  }\label{qj}
\end{figure}

The in-plane spin components can be  turned on  by applied out-of-plane  electric field,  which provides a possibility of the on-off logical functionality of spin devices by controlling  the passage of electrons (see \autoref{qj}). For example, two ferromagnetic (FM) electrodes in the same direction are set for  source and drain electrodes.  Without the out-of-plane electric
field, the out-of-plane orientation of the spin polarization injected from the source electrode maintains    due to PST, when the injected electrons pass through the T-SbP channel (\autoref{qj} (a)).
When the external electric field is applied,  the PST is broken, and  the PSH state is perturbed.
The spin of  electrons in the channel will  rotate, and will be blocked by the drain electrode (\autoref{qj} (b)).  The function of spintronic switch can be realized in the possible  spintronic device.

\section{Discussion and Conclusion}
A general form of the SOC Hamiltonian $H_{SOC}$  in solid-state materials can be expressed as\cite{q7-9,q7-10}:
 \begin{equation}\label{h11}
    H_{SOC}=\alpha_1(\vec{E}\times\vec{k})\cdot\vec{\sigma}
 \end{equation}
where $\alpha_1$ is the strength of the SOC, and $\vec{\sigma}$=($\sigma_x$, $\sigma_y$, $\sigma_z$) are the Pauli matrices.
For T-SbP, the spontaneous in-plane
electric polarization is oriented along $x$ direction.
The \autoref{h11} is written as:
\begin{equation}\label{h12}
    H_{SOC}=\alpha_1 E_x k_y\sigma_z
 \end{equation}
This is consistent with \autoref{h}, when ignoring  the higher order terms of $k$.

The electric field  along $x$ direction can also be induced with a uniaxial in-plane strain  by piezoelectric effect, and then tune  spin-splitting of T-SbP.
When a uniaxial in-plane vibration is applied to T-SbP, the spin-splitting will be modulated periodically. As shown in FIG.11 (a)\cite{bc}, the compressive strain produces negative electric field, and then reduces spin-splitting of PST. However, FIG.11 (b)\cite{bc} shows that the tensile strain induces positive electric field, and then enhances spin-splitting of PST. So, the  orthorhombic primitive cell  is used to calculate the   piezoelectric  stress  coefficients $e_{ij}$ of T-XY (X$\neq$Y=P, As, Sb and Bi), and then the  piezoelectric  strain  coefficients $d_{ij}$ are obtained by the $C_{ij}$ and $e_{ij}$\cite{apl}. The related data are listed in TABLE I\cite{bc}. The $|d_{11}|$ of T-SbP  is as high as 266.15 pm/V, which  plays a positive
effect on tuning spin-splitting by piezoelectric effect.
Piezotronic effect on Rashba SOC in a ZnO/P3HT nanowire array structure has been investigated experimentally\cite{ydt}. It is found that the Rashba SOC can be effectively tuned by inner-crystal piezo-potential created inside the ZnO nanowires instead of an externally applied voltage. So, the spin-splitting of PST in T-XY (X$\neq$Y=P, As, Sb and Bi) can be tuned by piezoelectric effect.

In summary, based on the first-principles calculations along with symmetry analysis,
we have systematically investigated the electronic  properties of  2D T-XY (X$\neq$Y=P, As and Sb) with dynamical, mechanical and thermal  stabilities.  Because of  $C_{2\upsilon}$ point-group symmetry, the unidirectional out-of-plane spin configurations
are preserved in 2D T-XY (X$\neq$Y=P, As and Sb), giving rise to the PSH state. It is found that  this PSH is observed near  both VBM and CBM around $\Gamma$ point. The out-of-plane electric field  can be used to perturb PSH, which provides possibility to realize electronic device of switching function.  Our works reveal a new 2D family of materials that  have great potential for spintronic  device applications.

\begin{acknowledgments}
This work was supported by Natural Science Basis Research Plan
in Shaanxi Province of China (No. 2021JM-456). We are grateful to
Shanxi Supercomputing Center of China, and the calculations were
performed on TianHe-2.
\end{acknowledgments}


\begin{references}

\bibitem{p1}E. I. Rashba, Sov. Phys. Solid State \textbf{2}, 1224 (1960).

\bibitem{p2}G. Dresselhaus, Phys. Rev. \textbf{100}, 580 (1955).

\bibitem{p3}J. Nitta, T. Akazaki, H. Takayanagi, and T. Enoki, Phys. Rev.
Lett. \textbf{78}, 1335 (1997).

\bibitem{p4} A. Manchon, H. C. Koo, J. Nitta, S. M. Frolov, and R. A. Duine,
Nat. Mater. \textbf{14}, 871 (2015).

\bibitem{p5} P. Chuang, S. H. Ho, L. W. Smith et al., Nat. Nanotechnol. \textbf{10}, 35
(2009).


\bibitem{p6}M. I. Dyakonov and V. I. Perel, Sov. Phys. Solid State \textbf{13}, 3023 (1972).


\bibitem{p7}B. A. Bernevig, J. Orenstein, and S.-C. Zhang, Phys. Rev. Lett.
\textbf{97}, 236601 (2006).

\bibitem{p8}J. Schliemann, Rev. Mod. Phys. \textbf{89}, 011001 (2017).

\bibitem{p9}P. Altmann, M. P. Walser, C. Reichl, W. Wegscheider, and G.
Salis, Phys. Rev. B \textbf{90}, 201306(R) (2014).


\bibitem{p10}J. D. Koralek, C. P. Weber, J. Orenstein, B. A. Bernevig, S.-C.
Zhang, S. Mack, and D. D. Awschalom, Nature (London)  \textbf{458},
610 (2009).
\bibitem{p11}M. P. Walser, C. Reichl, W. Wegscheider, and G. Salis, Nat.
Phys.  \textbf{8}, 757 (2012).


\bibitem{p12}J. Ishihara, Y. Ohno, and H. Ohno, Appl. Phys. Express  \textbf{7},
013001 (2014).

\bibitem{p13}M. Kohda, V. Lechner, Y. Kunihashi et al., Phys. Rev. B  \textbf{86},
081306(R) (2012).

\bibitem{p13-1}A. Sasaki, S. Nonaka, Y. Kunihashi, M. Kohda, T. Bauernfeind,
T. Dollinger, K. Richter, and J. Nitta, Nat. Nanotechnol.  \textbf{9}, 703
(2014).


\bibitem{p14}N. Yamaguchi and F. Ishii, Appl. Phys. Express  \textbf{10}, 123003
(2017).

\bibitem{p15}M. A. U. Absor, F. Ishii, H. Kotaka, and M. Saito, Appl. Phys.
Express  \textbf{8}, 073006 (2015).




\bibitem{p16}L. L. Tao  and E. Y. Tsymbal,  Nat. Commun. \textbf{9}, 2763 (2018).


\bibitem{p17}C. Autieri, P. Barone, J. Slawi$\acute{n}$ska  and S. Picozzi,  Phys. Rev. Materials \textbf{3}, 084416 (2019).


\bibitem{p18}H. Djani, A. C. Garcia-Castro, W. Y. Tong, P.  Barone, E. Bousquet, S. Picozzi  and P. Ghosez, npj Quant. Mater.  \textbf{4}, 51 (2019).


\bibitem{p19}H. Ai, X. Ma, X. Shao, W. Li  and M. Zhao,  Phys. Rev. Materials \textbf{3}, 054407 (2019).

\bibitem{p20}S. A. Sasmito, M. Anshory, I. Jihad  and M. A. U. Absor,  Phys. Rev. B \textbf{104}, 115145 (2021).
%Reversible spin textures with giant spin splitting in two-dimensional GaXY (X = Se, Te; Y = Cl, Br, I) compounds for a persistent spin helix
\bibitem{p21}M. A. U. Absor  and F. Ishii,  Phys. Rev. B \textbf{103}, 045119 (2021).


\bibitem{p22}F. Jia, S. Hu, S. Xu, H.  Gao, G. Zhao, P.  Barone, A.  Stroppa  and  W. Ren,   J. Phys. Chem. Lett. \textbf{11}, 5177 (2020).



\bibitem{p23}M. A. U. Absor and F. Ishii, Phys. Rev. B  \textbf{99}, 075136 (2019).


\bibitem{p24}M. A. U. Absor  and F. Ishii,  Phys. Rev. B \textbf{100}, 115104 (2019).
%Intrinsic persistent spin helix state in two-dimensional group-IV monochalcogenide MX monolayers (M = Sn or Ge and X = S, Se, or Te)

\bibitem{p25}H. Lee, J. Im  and H. Jin,  Appl. Phys. Lett. \textbf{116}, 022411 (2020).

\bibitem{p26}H. J. Zhao, H. Nakamura, R. Arras, C.  Paillard, P.  Chen, J. Gosteau, X. Li, Y. Yang  and L. Bellaiche, Phys. Rev. Lett. \textbf{125}, 216405 (2020).



\bibitem{apl}H. Lei, T. Ouyang, C. Y. He, J. Li and C. Tang, Appl. Phys. Lett. 122, 062903 (2023).



%%%%%%%%%%%%%%%%%%%%%%%%%%%%%%%%%%%%%%%%%%%%%%%%%%%%%%%%%%%%%%%%%%%%
\bibitem{1}P. Hohenberg and W. Kohn, Phys. Rev. \textbf{136},
B864 (1964); W. Kohn and L. J. Sham, Phys. Rev. \textbf{140},
A1133 (1965).

\bibitem{pv1} G. Kresse, J. Non-Cryst. Solids \textbf{193}, 222 (1995).

\bibitem{pv2} G. Kresse and J. Furthm$\ddot{u}$ller, Comput. Mater. Sci. 6, \textbf{15} (1996).

\bibitem{pv3} G. Kresse and D. Joubert, Phys. Rev. B \textbf{59}, 1758 (1999).

\bibitem{pbe}J. P. Perdew, K. Burke and M. Ernzerhof, Phys. Rev. Lett. \textbf{77}, 3865 (1996).

%\bibitem{u}S. L. Dudarev, G. A. Botton, S. Y. Savrasov, C. J. Humphreys and A. P. Sutton, Phys. Rev. B \textbf{57}, 1505 (1998).


\bibitem{soc} S. Steiner, S. Khmelevskyi, M. Marsmann and G. Kresse, Phys. Rev. B \textbf{93}, 224425 (2016).

\bibitem{pv5}A. Togo, F. Oba, and I. Tanaka, Phys. Rev. B \textbf{78}, 134106
(2008).



\bibitem{py}U. Herath, P. Tavadze, X. He, E. Bousquet, S. Singh, F. Munoz and A. H. Romero,  Comput. Phys. Commun. \textbf{251}, 107080 (2020).


%%%%%%%%%%%%%%%%%%%%%%%%%%%%%%%%%%%%%%%%%%%%%%%%%%%%%%%%%%%%%%%%%%%%%%%%%%%%%%%%%%%%%%%%%%%%%
\bibitem{bc}See Supplemental Material at [] for the phonon band dispersions, AIMD results and energy band structures of T-AsP, T-SbAs, T-BiP, T-BiAs  and T-BiSb; the  direction-dependent Young's modulus $C_\text{2D}(\theta)$ and   piezoelectric coefficients   $e_{ij}$/$d_{ij}$  of T-XY (X$\neq$Y=P, As, Sb and Bi);  the energy band structures  of T-SbP  as a function of electric field $E$; schematic of  strain-tuned  persistent spin-splitting.


\bibitem{ela}M. Born and K. Huang, Am. J. Phys. \textbf{23}, 474 (1995).


\bibitem{ela1}E. Cadelano, P. L. Palla, S. Giordano and L. Colombo,  Phys. Rev. B  \textbf{82}, 235414 (2010).

\bibitem{q7-9}J. Nitta, T. Akazaki, H. Takayanagi and T. Enoki, Phys. Rev.
Lett. \textbf{78}, 1335 (1997).

\bibitem{q7-10}A. Manchon, H. C. Koo, J. Nitta, S. M. Frolov and R. A. Duine,
Nat. Mater. \textbf{14}, 871 (2015).


\bibitem{ydt}L. Zhu, Y. Zhang, P. Lin et al., ACS Nano \textbf{12}, 1811 (2018).
\end{references}
\end{document}